\documentclass[11pt,a4paper,english,nofootinbib]{revtex4}
\usepackage{lmodern}
\usepackage{lmodern}

\usepackage[T1]{fontenc}
\usepackage[latin9]{inputenc}
\usepackage{babel}
\usepackage{amsmath}
\usepackage{amssymb}
\usepackage{esint}
\usepackage[unicode=true,pdfusetitle,
 bookmarks=true,bookmarksnumbered=false,bookmarksopen=false,
 breaklinks=false,pdfborder={0 0 1},backref=false,colorlinks=false]
 {hyperref}

\makeatletter

\@ifundefined{textcolor}{}
{%
 \definecolor{BLACK}{gray}{0}
 \definecolor{WHITE}{gray}{1}
 \definecolor{RED}{rgb}{1,0,0}
 \definecolor{GREEN}{rgb}{0,1,0}
 \definecolor{BLUE}{rgb}{0,0,1}
 \definecolor{CYAN}{cmyk}{1,0,0,0}
 \definecolor{MAGENTA}{cmyk}{0,1,0,0}
 \definecolor{YELLOW}{cmyk}{0,0,1,0}
}

\usepackage{babel}
\usepackage{babel}
\usepackage{babel}
\usepackage{babel}
\usepackage{babel}
\usepackage{babel}

\usepackage{babel}
\usepackage{graphicx}
\def\b{\begin{equation}}
\def\e{\end{equation}}

\@ifundefined{textcolor}{}{%
 \definecolor{BLACK}{gray}{0}
 \definecolor{WHITE}{gray}{1}
 \definecolor{RED}{rgb}{1,0,0}
 \definecolor{GREEN}{rgb}{0,1,0}
 \definecolor{BLUE}{rgb}{0,0,1}
 \definecolor{CYAN}{cmyk}{1,0,0,0}
 \definecolor{MAGENTA}{cmyk}{0,1,0,0}
 \definecolor{YELLOW}{cmyk}{0,0,1,0}
 }

\usepackage{latexsym}\usepackage{bm}

\makeatother

\begin{document}
\title{{\normalsize{}{}Energy and Angular Momentum in $D$ dimensional Kerr-AdS
black holes-new formulation}}
\author{{\normalsize{}{}Emel Altas}}
\email{emelaltas@kmu.edu.tr}

\affiliation{Department of Physics,\\
 Karamanoglu Mehmetbey University, 70100, Karaman, Turkey}
\date{{\normalsize{}{}\today}}
\begin{abstract}
Recently it was shown that the conserved charges of asymptotically
anti de Sitter spacetimes can be written in an explicitly gauge-invariant
way in terms of the linearized Riemann tensor and the Killing vectors.
Here we employ this construction to compute the mass and angular momenta
of the $D$ dimensional Kerr-AdS black holes, which is one of the
most remarkable Einstein metrics generalizing the four dimensional
rotating black hole. 
\end{abstract}
\maketitle

\section{Introduction}

Recently, we have given a new formulation of conserved charges in
cosmological Einstein's theory \citep{newformula} and quadratic theories
\citep{newformulauzun} which is explicitly gauge invariant. The formula
for asymptotically anti de Sitter (AdS) spacetimes reads
\begin{equation}
Q=\frac{(D-1)(D-2)}{8G_{D}\Omega_{D-2}\Lambda\left(D-3\right)}\int_{\partial\bar{\Sigma}}dS_{r}\left(R^{r0}\thinspace_{\beta\sigma}\right)^{\left(1\right)}\bar{\nabla}^{\beta}\bar{\xi}^{\sigma}.\label{newcharges}
\end{equation}
Here the integral is over the boundary of the spatial hypersurface
$\varSigma$, $\left(R^{r0}\thinspace_{\beta\sigma}\right)^{\left(1\right)}$
is the linearized Riemann tensor, $\bar{\xi}^{\mu}$ is a background
Killing vector of the AdS spacetime and $G_{D}$ denotes the Newton's
constant, $\Omega_{D-2}$ is the solid angle. The relation between
this formula and the Abbott-Deser \citep{AD} formula, which is gauge
invariant up to a boundary term, was given in \citep{newformula}.
In \citep{newformula} the new formula was used to compute the conserved
mass and angular momentum of the four dimensional Kerr-AdS black hole
and the results are consistent with the other methods \citep{deserkaniktekin}.
Here we extend the discussion to generic $D$ dimensions for which
the computation is much more complex. The relevant solution, that
is the $D$ dimensional Kerr-AdS metric, was quite hard to find and
in fact it was constructed in $2004$ in \citep{gibbons1,Gibbons_2005}.
Conserved charges of this metric was computed with various techniques
including the AD technique \citep{deserkaniktekin}. Here our task
is twofold: we shall give a computation of the conserved charges for
these metrics with the new formula and provide all the relevant details
of the computations which are missing in the previous literature.
But before that, let us recap some work on conserved charges in gravity
theories.

Construction of the conserved quantities has picked up interest for
various spacetimes. For an asymptotically flat spacetime one has the
Arnowitt-Deser-Misner (ADM) mass \citep{ADM}, which yields exactly
the expected energy of an isolated gravitational system. For asymptotically
AdS spacetimes, Abbott and Deser generalized the ADM mass and they
constructed the Abbott-Deser (AD) charges \citep{AD} in cosmological
Einstein's gravity. Another generalization of the conserved charges
is the Abbott-Deser-Tekin formulation (ADT) \citep{AD,DT}, which
gives the conserved mass and angular momentum for AdS spacetimes where
higher curvature terms generically bring nontrivial contributions
to the charges. A detailed review of these constructions and many
applications has been recently given in \citep{sisman-tekin-setare}.

The layout of the paper is as follows: in section $\text{\mbox{II}}$,
we construct the energy and angular momentum of the $D$ dimensional
Kerr AdS metric solutions in cosmological Einstein's gravity using
the charge expression given in new formulation (\ref{newcharges}).
Some of the computations are relegated to the Appendices.

\section{Kerr-AdS Black Holes in D dimensions}

The mass and angular momentum of the $D$ dimensional Kerr-AdS black
holes are constructed in \citep{deserkaniktekin} using the AD formulation
\citep{AD,DT}. The conserved charges was constructed for the four
dimensional Kerr-AdS black holes in \citep{newformula} using the
new formula (\ref{newcharges}). Here, we extend the discussion and
compute the conserved charges of the $D$-dimensional Kerr-AdS black
holes for asymptotically (anti) de Sitter spacetimes in cosmological
Einstein's gravity. We use the metric given in \citep{gibbons1},
which is in the Kerr-Schild form\footnote{For the exact solutions of Einstein's theory and the importance of
the Kerr-Schild ansatz see \citep{TezzEtt}.} \citep{kerrschild1,gursey} as 
\begin{equation}
{\rm d}s^{2}={\rm d}\bar{s}^{2}+\frac{2MG_{D}}{U}\left(k_{\mu}{\rm d}x^{\mu}\right)^{2},\label{eq:KS}
\end{equation}
where $M$ is a real parameter and $U$ is defined as follows

\begin{equation}
U:=r^{\epsilon}\sum_{i=1}^{N+\epsilon}\frac{\mu_{i}^{2}}{r^{2}+a_{i}^{2}}\prod_{j=1}^{N}\left(r^{2}+a_{j}^{2}\right).
\end{equation}
Although, the Kerr-Schild metrics are the exact solutions of the cosmological
Einstein's gravity, one can use the perturbation theory expressing
the first order expansion\footnote{The inverse metric takes the simple form, $g^{\mu\nu}=\overline{g}^{\mu\nu}-h^{\mu\nu}$.
The higher order variations of the metric do not survive due to nullity
of $k_{\mu}$.} of the metric tensor as 
\begin{equation}
h_{\mu\nu}=\frac{2MG_{D}}{U}k_{\mu}k_{\nu}.\label{h_munu}
\end{equation}
The one form $k_{\mu}$ is null both with respect to the metrics $g$
and $\overline{g}$
\begin{equation}
k_{\mu}k_{\nu}g^{\mu\nu}=0=k_{\mu}k_{\nu}\overline{g}^{\mu\nu},
\end{equation}
 and it is given with 
\begin{equation}
k_{\mu}{\rm d}x^{\mu}=F{\rm d}r+W{\rm d}t-\sum_{i=1}^{N}\frac{a_{i}\mu_{i}^{2}}{1+\Lambda a_{i}^{2}}{\rm d}\phi_{i}.\label{k_mu's}
\end{equation}
One has the constraint equation $\sum_{i=1}^{N+\epsilon}\mu_{i}^{2}=1$,
where $\epsilon=1,~N=\frac{D-2}{2}$ for even dimensions and $\epsilon=0,~N=\frac{D-1}{2}$
for odd dimensions. The $N$ shows the number of rotation parameters.
The background metric in (\ref{eq:KS}) is the following de Sitter
metric
\begin{align}
{\rm d}\bar{s}^{2}= & -W\left(1-\Lambda r^{2}\right){\rm d}t^{2}+F{\rm d}r^{2}+\sum_{i=1}^{N+\epsilon}\frac{r^{2}+a_{i}^{2}}{1+\Lambda a_{i}^{2}}{\rm d}\mu_{i}^{2}+\sum_{i=1}^{N}\frac{r^{2}+a_{i}^{2}}{1+\Lambda a_{i}^{2}}\mu_{i}^{2}{\rm d}\phi_{i}^{2}\nonumber \\
 & +\frac{\Lambda}{W\left(1-\Lambda r^{2}\right)}\left(\sum_{i=1}^{N+\epsilon}\frac{\left(r^{2}+a_{i}^{2}\right)\mu_{i}{\rm d}\mu_{i}}{1+\Lambda a_{i}^{2}}\right)^{2},\label{eq:dS}
\end{align}
where $W$ and $F$ are defined as
\begin{equation}
W:=\sum_{i=1}^{N+\epsilon}\frac{\mu_{i}^{2}}{1+\Lambda a_{i}^{2}},\qquad\qquad\qquad F:=\frac{1}{1+\Lambda a_{i}^{2}}\sum_{i=1}^{N+\epsilon}\frac{r^{2}\mu_{i}^{2}}{r^{2}+a_{i}^{2}}.\label{definition-W-F}
\end{equation}
For even dimensional case $a_{N+1}=0$, since the $\phi_{N+1}$ component
is missing. The background metric (\ref{eq:dS}), yields the following
expressions
\begin{equation}
\bar{R}_{\mu\alpha\nu\beta}=\Lambda\left(\bar{g}_{\mu\nu}\bar{g}_{\alpha\beta}-\bar{g}_{\mu\beta}\bar{g}_{\alpha\nu}\right),\qquad\bar{R}_{\mu\nu}=\Lambda(D-1)\bar{g}_{\mu\nu},\qquad\bar{R}=\Lambda D(D-1).\label{dsriemann-ricci}
\end{equation}
The charge expression (\ref{newcharges}), is constructed for (anti)
de Sitter background metrics. So, we need to reexpress the conserved
charges by rescaling the cosmological constant $\Lambda$ as $\frac{(D-1)(D-2)\Lambda}{2}$.
The conserved charge expression then becomes
\begin{equation}
Q=\frac{1}{4G_{D}\Omega_{D-2}\Lambda\left(D-3\right)}\int_{\partial\bar{\Sigma}}dS_{r}\thinspace\left(R^{r0}\thinspace_{\beta\sigma}\right)^{\left(1\right)}\bar{\nabla}^{\beta}\bar{\xi}^{\sigma}.\label{chargeformulaforkerrschildform}
\end{equation}
Now, we can calculate the energy and angular momentum of the solutions
given in (\ref{eq:KS}). For the energy, we use the energy Killing
vector, $\bar{\xi}^{\mu}=(-1,\vec{0})$. Let us compute the integrand
in the last equation. Using the symmetries of the Riemann tensor and
antisymmetry of the Killing equation, one has
\begin{equation}
\left(R^{r0}\thinspace_{\beta\sigma}\right)^{\left(1\right)}\bar{\nabla}^{\beta}\bar{\xi}^{\sigma}=2\overline{g}^{00}\bar{\nabla}^{\beta}\bar{\xi}^{\sigma}\overline{\nabla}_{\beta}(\Gamma_{\rho\sigma}^{r})^{(1)}-\overline{R}^{r}\thinspace_{\rho\beta\sigma}h^{\rho0}\bar{\nabla}^{\beta}\bar{\xi}^{\sigma},
\end{equation}
where the covariant derivative of the Killing vector yields
\begin{equation}
\bar{\nabla}^{\beta}\bar{\xi}^{\sigma}=\frac{1}{2}\overline{g}^{\beta\nu}\overline{g}^{\sigma\gamma}(\partial_{\gamma}\overline{g}_{\nu0}-\partial_{\nu}\overline{g}_{\gamma0}).
\end{equation}
Then, we can express the integrand as
\begin{equation}
\left(R^{r0}\thinspace_{\beta\sigma}\right)^{\left(1\right)}\bar{\nabla}^{\beta}\bar{\xi}^{\sigma}=(\overline{g}^{00})^{2}\overline{g}^{\beta\nu}\partial_{\nu}\overline{g}_{00}\left(\overline{\nabla}_{0}(\Gamma_{0\beta}^{r})^{(1)}-\overline{\nabla}_{\beta}(\Gamma_{00}^{r})^{(1)}\right)+\partial_{\nu}\overline{g}_{00}\overline{R}^{r}\thinspace_{\rho}\thinspace^{\nu0}h^{\rho0}.
\end{equation}
The first order perturbation of the Christoffel symbol reads
\begin{equation}
(\Gamma_{\rho\sigma}^{r})^{(1)}=\frac{1}{2}(\overline{\nabla}h_{\sigma}^{r}+\overline{\nabla}_{\sigma}h_{\rho}^{r}-\overline{\nabla}^{r}h_{\rho\sigma}).
\end{equation}
So, one obtains
\begin{align}
 & \left(R^{r0}\thinspace_{\beta\sigma}\right)^{\left(1\right)}\bar{\nabla}^{\beta}\bar{\xi}^{\sigma}=\frac{1}{2}\partial_{\nu}\overline{g}_{00}\left(\overline{R}^{0\nu0\rho}h_{\rho}^{r}-\overline{R}^{0\nu r\rho}h_{\rho}^{0}\right)\label{khgv}\\
 & ~~~~~~~~~~~~~~~~~~~~~+\frac{1}{2}(\overline{g}^{00})^{2}\overline{g}^{rr}\overline{g}^{\beta\nu}\partial_{\nu}\overline{g}_{00}\left(\overline{\nabla}_{0}\overline{\nabla}_{0}h_{\beta r}+\overline{\nabla}_{\beta}\overline{\nabla}_{r}h_{00}-\overline{\nabla}_{0}\overline{\nabla}_{r}h_{\beta0}-\overline{\nabla}_{\beta}\overline{\nabla}_{0}h_{0r}\right),\nonumber 
\end{align}
where the components of the background metric are the functions of
both the $r$ and $\mu_{i}$ components. Hence, in the last equation
only $\nu=r$ and $\nu=\mu_{i}$ survive. One finds\footnote{See Appendix C for the details of the construction.},
\begin{equation}
\left(R^{r0}\thinspace_{\beta\sigma}\right)^{\left(1\right)}\bar{\nabla}^{\beta}\bar{\xi}^{\sigma}=2M\Lambda r^{2-D}(D-3)\Bigl(W(D-1)-1\Bigr).
\end{equation}
Substituting the integrand in (\ref{chargeformulaforkerrschildform}),
one arrives at
\begin{equation}
Q=\frac{M}{2G_{D}\Omega_{D-2}}\int_{\partial\bar{\Sigma}}dS_{r}\,r^{2-D}\Bigl(W(D-1)-1\Bigr).
\end{equation}
Defining the following functions
\begin{equation}
\Xi\equiv\prod_{i=1}^{N}\left(1+\Lambda a_{i}^{2}\right),~~~~~~~~~~~~~~~~~~~~~~~~\Xi_{i}\equiv1+\Lambda a_{i}^{2},\label{=00005CXi}
\end{equation}
one ends up with the energy of the even dimensional Kerr-AdS black
holes as
\begin{equation}
E=\frac{M}{\Xi}\sum_{i=1}^{N}\frac{1}{\Xi_{i}}.
\end{equation}
In the general case, the energy of the $D$-dimensional rotating black
holes can be written as 
\begin{equation}
E=\frac{M}{\Xi}\sum_{i=1}^{N}\left(\frac{1}{\Xi_{i}}-\frac{1}{2}\left(1-\epsilon\right)\right),\label{eq:finalen-1}
\end{equation}
which matches with the result given in \citep{deserkaniktekin}. Similarly,
considering the Killing vector $\xi_{(i)}^{\mu}=\left(0,...,0,1_{i},0,...,0\right)$,
where $i$ refers to the $i^{\mbox{th}}$ azimuthal angle $\phi_{i}$,
one has
\begin{equation}
\overline{\nabla}^{\beta}\overline{\xi}^{\sigma}=\frac{1}{2}(\overline{g}^{\beta\nu}\overline{g}^{\sigma\phi_{j}}-\overline{g}^{\beta\phi_{j}}\overline{g}^{\sigma\nu})\partial_{\nu}\overline{g}_{\phi_{i}\phi_{j}}.
\end{equation}
Then the integrand becomes
\begin{eqnarray}
 &  & \left(R^{r0}\thinspace_{\beta\sigma}\right)^{\left(1\right)}\bar{\nabla}^{\beta}\bar{\xi}^{\sigma}=\frac{1}{2}\partial_{\nu}\overline{g}_{\phi_{i}\phi_{j}}\left(\overline{R}^{\nu\phi_{j}0\rho}h_{\rho}^{r}-\overline{R}^{\nu\phi_{j}r\rho}h_{\rho}^{0}\right)\\
 &  & ~~~~~~~~~~~~~~~~~~~+\frac{1}{2}\overline{g}^{rr}\overline{g}^{00}\overline{g}^{\phi_{k}\phi_{j}}\overline{g}^{\nu\beta}\partial_{\nu}\overline{g}_{\phi_{i}\phi_{j}}\left(\overline{\nabla}_{\beta}\overline{\nabla}_{0}h_{\phi_{k}r}+\overline{\nabla}_{\phi_{k}}\overline{\nabla}_{r}h_{\beta0}-\overline{\nabla}_{\beta}\overline{\nabla}_{r}h_{\phi_{k}0}-\overline{\nabla}_{\phi_{k}}\overline{\nabla}_{0}h_{\beta r}\right).\nonumber 
\end{eqnarray}
Expressing the covariant derivatives in terms of the partial ones,
and computing the corresponding quantities one ends up with 
\begin{equation}
\left(R^{r0}\thinspace_{\beta\sigma}\right)^{\left(1\right)}\bar{\nabla}^{\beta}\bar{\xi}^{\sigma}=2M\Lambda r^{2-D}(D-3)(D-1)\frac{a_{i}\mu_{i}^{2}}{1+\Lambda a_{i}^{2}}.
\end{equation}
Then, the angular momentum can be expressed as
\begin{equation}
J_{i}=\frac{M(D-1)}{2G_{D}\Omega_{D-2}}\int_{\partial\bar{\Sigma}}dS_{r}\thinspace r^{2-D}\frac{a_{i}\mu_{i}^{2}}{1+\Lambda a_{i}^{2}}.
\end{equation}
Using the definitions in (\ref{=00005CXi}), one arrives at the angular
momentum of the even dimensional Kerr-AdS black holes as 
\begin{equation}
J_{i}=\frac{Ma_{i}}{\Xi\Xi_{i}},
\end{equation}
which is same with the expression given in \citep{deserkaniktekin}.
One can express the total energy in terms of angular momenta as follows
\begin{equation}
E=\sum_{i=1}^{N}\frac{J_{i}}{a_{i}},
\end{equation}
in even dimensions. Also, one has the relation given below
\begin{equation}
E=\sum_{i=1}^{N}\frac{J_{i}}{a_{i}}-\frac{NM}{2\Xi}
\end{equation}
between the energy and angular momenta for the metric solutions (\ref{eq:KS})
of cosmological Einstein's gravity in odd dimensional case.

\section{CONCLUSIONS}

In cosmological Einstein's theory the conserved charges are explicitly
gauge-invariant, under small diffeomorphisms generated by a background
vector field, for asymptotically AdS spacetimes. But the current expression,
which yields the conserved energy and angular momentum, may be gauge-invariant
up to a boundary term. The Abbott-Deser (AD) \citep{AD} formulation
of the conserved charges obtained from such a current two form, which
is not explicitly gauge-invariant. Recently, we have given a new method
to construct the conserved charges \citep{newformula}, in which the
starting point is the second Bianchi identity instead of the explicit
form of the cosmological Einstein tensor in terms of the metric tensor.
Also, we have shown that it is possible to use this formulation to
construct the conserved charges in generic gravity theories \citep{newformulauzun}.
In this new formulation, the resulting charge expression involves
the linearized Riemann tensor and not only the conserved charges but
also the current expression is explicitly gauge-invariant. 

The $D$ dimensional Kerr-AdS metric solutions of cosmological Einstein's
gravity are in a complex form which makes the calculation of the conserved
charges so difficult. But these solutions are too important since
they represent the rotating black holes of the theory for the generic
$D$ dimensions. In \citep{deserkaniktekin} the conserved charges
of the $D$ dimensional Kerr-AdS black holes was constructed with
the Abbott-Deser (AD) \citep{AD} formulation. In \citep{newformula}
we computed the mass and angular momentum of the four dimensional
Kerr- AdS black holes with the new energy formula, which yields the
matching results with \citep{deserkaniktekin}. In this paper, we
extend the discussion and construct the conserved charges of the $D$
dimensional Kerr-AdS black holes using the new formulation. The expressions
we obtained for the $D$ dimensional case are consistent with the
results given in \citep{deserkaniktekin}.
\begin{acknowledgments}
The Author would like to thank Prof. Dr. Bayram Tekin for his helpful
remarks and discussion on conserved charges in cosmological Einstein's
gravity.
\end{acknowledgments}

\section*{APPENDIX A: DETERMINANT OF THE BACKGROUND METRIC TENSOR }

In this part, we compute the determinant of the background metric,
de Sitter metric in $D$ spacetime dimensions (\ref{eq:dS}), assuming
$D$ is even. One can express 
\begin{equation}
\det\overline{g}_{\mu\nu}=W(\Lambda r^{2}-1)F\prod_{i=1}^{N}\frac{r^{2}+a_{i}^{2}}{1+\Lambda a_{i}^{2}}\mu_{i}^{2}\det\overline{g}_{\mu_{i}\mu_{j}}.
\end{equation}
Since we need the $r\rightarrow\infty$ limit of the determinant,
we can express the last equation as
\begin{equation}
\det\overline{g}_{\mu\nu}=-Wr^{2N}\prod_{i=1}^{N}\frac{\mu_{i}^{2}}{1+\Lambda a_{i}^{2}}det\overline{g}_{\mu_{i}\mu_{j}}.\label{largelimitofthedeterminant}
\end{equation}
Here we used the large $r$ limit of the $F$ function, defined in
(\ref{definition-W-F}), which explicitly reads $F=-1/\Lambda r^{2}$.
Obviously, one needs to calculate the determinant of the non-diagonal
piece of the background metric. First, let us compute the $\overline{g}_{\mu_{i}\mu_{j}}$
component for the even dimensional case. The non-diagonal piece in
(\ref{eq:dS}) involves two terms. Let us compute these terms explicitly
and then collect the pieces. We have
\begin{equation}
\sum_{i=1}^{N+1}\frac{r^{2}+a_{i}^{2}}{1+\Lambda a_{i}^{2}}d\mu_{i}^{2}=\sum_{i=1}^{N}\frac{r^{2}+a_{i}^{2}}{1+\Lambda a_{i}^{2}}d\mu_{i}^{2}+\frac{r^{2}+a_{N+1}^{2}}{1+\Lambda a_{N+1}^{2}}d\mu_{N+1}^{2},
\end{equation}
where $a_{N+1}=0$. Then, the last equation reduces to 
\begin{equation}
\sum_{i=1}^{N+1}\frac{r^{2}+a_{i}^{2}}{1+\Lambda a_{i}^{2}}d\mu_{i}^{2}=\sum_{i=1}^{N}\frac{r^{2}+a_{i}^{2}}{1+\Lambda a_{i}^{2}}d\mu_{i}^{2}+r^{2}d\mu_{N+1}^{2}.\label{pieceoffdiagonal}
\end{equation}
 The constraint equation $\sum_{i=1}^{N+\epsilon}\mu_{i}^{2}=1$,
can be rewritten as follows 
\begin{equation}
\mu_{N+1}^{2}=1-\sum_{i=1}^{N}\mu_{i}^{2}.
\end{equation}
Taking the derivative of the last expression, one obtains
\begin{equation}
d\mu_{N+1}=-\frac{\sum_{i=1}^{N}\mu_{i}d\mu_{i}}{1-\sum_{i=1}^{N}\mu_{i}^{2}},\label{derivativeofconstraint}
\end{equation}
and equation (\ref{pieceoffdiagonal}) becomes
\begin{equation}
\sum_{i=1}^{N+1}\frac{r^{2}+a_{i}^{2}}{1+\Lambda a_{i}^{2}}d\mu_{i}^{2}=\sum_{i=1}^{N}\frac{r^{2}+a_{i}^{2}}{1+\Lambda a_{i}^{2}}d\mu_{i}^{2}-\frac{r^{2}}{\mu_{N+1}^{2}}\sum_{i=1}^{N}\sum_{j=1}^{N}\mu_{i}\mu_{j}d\mu_{i}d\mu_{j}.
\end{equation}
Similarly, we can express the remaining non-diagonal piece in (\ref{eq:dS})
as
\begin{equation}
\Bigl(\sum_{i=1}^{N+1}\frac{r^{2}+a_{i}^{2}}{1+\Lambda a_{i}^{2}}\mu_{i}d\mu_{i}\Bigr)^{2}=\Bigl(\sum_{i=1}^{N}\frac{r^{2}+a_{i}^{2}}{1+\Lambda a_{i}^{2}}\mu_{i}d\mu_{i}-r^{2}\mu_{i}d\mu_{i}\Bigr)^{2},
\end{equation}
where we have used the derivative of the constraint equation (\ref{derivativeofconstraint}),
and $a_{N+1}=0$. We can express the last equation as
\begin{equation}
\Bigl(\sum_{i=1}^{N+1}\frac{r^{2}+a_{i}^{2}}{1+\Lambda a_{i}^{2}}\mu_{i}d\mu_{i}\Bigr)^{2}=\sum_{i=1}^{N}\sum_{j=1}^{N}\mu_{i}\mu_{j}d\mu_{i}d\mu_{j}\Bigl(\frac{r^{2}+a_{i}^{2}}{1+\Lambda a_{i}^{2}}-r^{2}\Bigr)\Bigl(\frac{r^{2}+a_{j}^{2}}{1+\Lambda a_{j}^{2}}-r^{2}\Bigr).
\end{equation}
Collecting the pieces, the $\overline{g}_{\mu_{i}\mu_{j}}$ component
reads
\begin{equation}
\overline{g}_{\mu_{i}\mu_{j}}=\frac{r^{2}+a_{i}^{2}}{1+\Lambda a_{i}^{2}}\delta_{ij}+\mu_{i}\mu_{j}\Bigl(\frac{r^{2}}{\mu_{N+1}^{2}}+\frac{\Lambda}{W(1-\Lambda r^{2})}\Bigl(\frac{r^{2}+a_{i}^{2}}{1+\Lambda a_{i}^{2}}-r^{2}\Bigr)\Bigl(\frac{r^{2}+a_{j}^{2}}{1+\Lambda a_{j}^{2}}-r^{2}\Bigr)\Bigr).
\end{equation}
Now, let us consider the $r\rightarrow\infty$ limit of the $\overline{g}_{\mu_{i}\mu_{j}}$.
One ends up with
\begin{equation}
\overline{g}_{\mu_{i}\mu_{j}}=\frac{r^{2}\delta_{ij}}{1+\Lambda a_{i}^{2}}+\frac{r^{2}\mu_{i}\mu_{j}}{\mu_{N+1}^{2}}-\frac{\Lambda^{2}r^{2}}{W}\frac{a_{i}^{2}a_{j}^{2}\mu_{i}\mu_{j}}{(1+\Lambda a_{i}^{2})(1+\Lambda a_{j}^{2})}.
\end{equation}
Note that $i,j,...=1,2,...N$, where $N=\frac{D-2}{2}$, and there
is no summation over the indices. For simplicity, let us express the
result in a compact form as

\begin{equation}
\overline{g}_{\mu_{i}\mu_{j}}=A_{i}\delta_{ij}+B_{i}B_{j}+C_{i}C_{j},\label{g-nondiagonal-short}
\end{equation}
where $A_{i},B_{i}$ and $C_{i}$ functions are defined as follows
\begin{equation}
A_{i}:=\frac{r^{2}}{1+\Lambda a_{i}^{2}},~~~~~~~~~B_{i}:=\frac{r\mu_{i}}{\mu_{N+1}},~~~~~~~C_{i}:=\frac{\Lambda r}{\sqrt{-W}}\frac{\mu_{i}a_{i}^{2}}{(1+\Lambda a_{i}^{2})}.\label{definitionaibici}
\end{equation}
Computing the determinant of a metric, which is in the form (\ref{g-nondiagonal-short})
is not so easy. To simplify the calculation, let us consider the lower
dimensional cases and then generalize the results. Start with the
$N=2$ case, we can express the determinant as
\begin{equation}
\det\overline{g}_{\mu_{i}\mu_{j}}=A_{1}A_{2}\Bigl(1+\frac{B_{1}^{2}}{A_{1}}+\frac{B_{2}^{2}}{A_{2}}+\frac{C_{1}^{2}}{A_{1}}+\frac{C_{2}^{2}}{A_{2}}+\frac{1}{A_{1}A_{2}}(B_{1}^{2}C_{2}^{2}+B_{2}^{2}C_{1}^{2}-2B_{1}B_{2}C_{1}C_{2})\Bigr).
\end{equation}
Note that $N=2$ corresponds to the six dimensional spacetime. When
$N=3$, one has $D=8$ and the determinant yields
\begin{eqnarray}
 &  & \det\overline{g}_{\mu_{i}\mu_{j}}=A_{1}A_{2}A_{3}\Bigl(1+\frac{B_{1}^{2}}{A_{1}}+\frac{B_{2}^{2}}{A_{2}}+\frac{B_{3}^{2}}{A_{3}}+\frac{C_{1}^{2}}{A_{1}}+\frac{C_{2}^{2}}{A_{2}}+\frac{C_{3}^{2}}{A_{3}}\nonumber \\
 &  & ~~~~~~~~+\frac{1}{A_{1}A_{2}}(B_{1}^{2}C_{2}^{2}+B_{2}^{2}C_{1}^{2}-2B_{1}B_{2}C_{1}C_{2})+\frac{1}{A_{1}A_{3}}(B_{1}^{2}C_{3}^{2}+B_{3}^{2}C_{1}^{2}-2B_{1}B_{3}C_{1}C_{3})\nonumber \\
 &  & ~~~~~~~~+\frac{1}{A_{2}A_{3}}(B_{2}^{2}C_{3}^{2}+B_{3}^{2}C_{2}^{2}-2B_{2}B_{3}C_{2}C_{3})\Bigr).
\end{eqnarray}
Then, we can generalize the results as follows
\begin{equation}
\det\overline{g}_{\mu_{i}\mu_{j}}=\prod_{k=1}^{N}A_{k}\Bigl(1+\sum_{j=1}^{N}\Bigl(\frac{B_{j}^{2}}{A_{j}}+\frac{C_{j}^{2}}{A_{j}}\Bigr)+\sum_{i=1}^{N}\sum_{j\neq i}^{N}\Bigl(\frac{B_{i}^{2}C_{j}^{2}}{A_{i}A_{j}}-\frac{B_{i}B_{j}C_{i}C_{j}}{A_{i}A_{j}}\Bigr)\Bigr).\label{determinantformula}
\end{equation}
Now, we need to express this result in terms of the $r$ and $\mu_{i}$
components. Inserting the functions given in (\ref{definitionaibici}),
one obtains
\begin{equation}
\frac{B_{j}^{2}}{A_{j}}+\frac{C_{j}^{2}}{A_{j}}=\frac{1}{W\mu_{N+1}^{2}}\Bigl(W\mu_{j}^{2}(1+\Lambda a_{j}^{2})-\Lambda^{2}\mu_{N+1}^{2}\frac{\mu_{j}^{2}a_{j}^{4}}{1+\Lambda a_{j}^{2}}\Bigr).\label{a}
\end{equation}
Also, the last two terms in (\ref{determinantformula}) yield the
following
\begin{equation}
\frac{B_{i}^{2}C_{j}^{2}}{A_{i}A_{j}}-\frac{B_{i}B_{j}C_{i}C_{j}}{A_{i}A_{j}}=\frac{\Lambda^{2}\mu_{i}^{2}\mu_{j}^{2}a_{j}^{2}(a_{i}^{2}-a_{j}^{2})}{W\mu_{N+1}^{2}(1+\Lambda a_{j}^{2})}.\label{b}
\end{equation}
Using the equations (\ref{a}, \ref{b}), we arrive at
\begin{eqnarray}
 &  & 1+\sum_{j=1}^{N}\Bigl(\frac{B_{j}^{2}}{A_{j}}+\frac{C_{j}^{2}}{A_{j}}\Bigr)+\sum_{i=1}^{N}\sum_{j\neq i}^{N}\Bigl(\frac{B_{i}^{2}C_{j}^{2}}{A_{i}A_{j}}-\frac{B_{i}B_{j}C_{i}C_{j}}{A_{i}A_{j}}\Bigr)=\frac{1}{W\mu_{N+1}^{2}}\Bigl(W\mu_{N+1}^{2}\label{hhgjk}\\
 &  & +W\sum_{j=1}^{N}\mu_{j}^{2}(1+\Lambda a_{j}^{2})-\Lambda^{2}\mu_{N+1}^{2}\sum_{j=1}^{N}\frac{\mu_{j}^{2}a_{j}^{4}}{1+\Lambda a_{j}^{2}}+\Lambda^{2}\sum_{i=1}^{N}\sum_{j\neq i}^{N}\frac{\mu_{i}^{2}\mu_{j}^{2}a_{j}^{2}(a_{i}^{2}-a_{j}^{2})}{W\mu_{N+1}^{2}(1+\Lambda a_{j}^{2})}\Bigr).\nonumber 
\end{eqnarray}
One can get rid of the $\mu_{N+1}^{2}$ terms, substituting the constraint
$\mu_{N+1}^{2}=1-\sum_{i=1}^{N}\mu_{i}^{2}$ in the last equation.
One obtains
\begin{eqnarray}
 &  & 1+\sum_{j=1}^{N}\Bigl(\frac{B_{j}^{2}}{A_{j}}+\frac{C_{j}^{2}}{A_{j}}\Bigr)+\sum_{i=1}^{N}\sum_{j\neq i}^{N}\Bigl(\frac{B_{i}^{2}C_{j}^{2}}{A_{i}A_{j}}-\frac{B_{i}B_{j}C_{i}C_{j}}{A_{i}A_{j}}\Bigr)=\frac{1}{W\mu_{N+1}^{2}}\Bigl(W+W\sum_{j=1}^{N}\Lambda\mu_{j}^{2}a_{j}^{2}\nonumber \\
 &  & ~~~~~~~~~~~~~~~~~~~~~~~~~~~~-\Lambda^{2}\sum_{j=1}^{N}\frac{\mu_{j}^{2}a_{j}^{4}}{1+\Lambda a_{j}^{2}}+\Lambda^{2}\sum_{i=1}^{N}\sum_{j\neq i}^{N}\frac{\mu_{i}^{2}\mu_{j}^{2}a_{j}^{2}(a_{i}^{2}-a_{j}^{2})}{W\mu_{N+1}^{2}(1+\Lambda a_{j}^{2})}\Bigr),\label{abc}
\end{eqnarray}
which can be reduced further expressing the function $W$, defined
in equation (\ref{definition-W-F}), as
\begin{equation}
W=1-\sum_{i=1}^{N}\frac{\Lambda\mu_{i}^{2}a_{i}^{2}}{1+\Lambda a_{i}^{2}}.
\end{equation}
Inserting $W$ in (\ref{abc}), one ends up with 
\begin{equation}
1+\sum_{j=1}^{N}\Bigl(\frac{B_{j}^{2}}{A_{j}}+\frac{C_{j}^{2}}{A_{j}}\Bigr)+\sum_{i=1}^{N}\sum_{j\neq i}^{N}\Bigl(\frac{B_{i}^{2}C_{j}^{2}}{A_{i}A_{j}}-\frac{B_{i}B_{j}C_{i}C_{j}}{A_{i}A_{j}}\Bigr)=\frac{1}{W\mu_{N+1}^{2}}.
\end{equation}
From (\ref{determinantformula}), one arrives the determinant of the
$\overline{g}_{\mu_{i}\mu_{j}}$ as 
\begin{equation}
\det\overline{g}_{\mu_{i}\mu_{j}}=\frac{1}{W\mu_{N+1}^{2}}\prod_{i=1}^{N}A_{i}=\frac{r^{2N}}{W\mu_{N+1}^{2}}\prod_{i=1}^{N}\frac{1}{1+\Lambda a_{i}^{2}}.\label{determinantofnondiagonalpiece}
\end{equation}
As a final step, we need to insert this expression in (\ref{largelimitofthedeterminant}).
We end up with
\begin{equation}
\det\overline{g}_{\mu\nu}=-\Bigl(\frac{r^{2N}}{\mu_{N+1}}\prod_{i=1}^{N}\frac{\mu_{i}}{1+\Lambda a_{i}^{2}}\Bigr)^{2},
\end{equation}
which yields
\begin{equation}
\sqrt{-\overline{g}}=\frac{r^{2N}}{\mu_{N+1}}\prod_{i=1}^{N}\frac{\mu_{i}}{1+\Lambda a_{i}^{2}}\label{determinantlast}
\end{equation}
for the even dimensional case. For the odd dimensional case, one can
carry out the similar calculation.

\section*{APPENDIX B: INTEGRAL EXPRESSIONS}

Here, we compute the integral expressions, that we have used in the
construction of the conserved charges of the $D$ dimensional Kerr-
AdS metrics. Let us calculate the integral given below
\begin{equation}
I_{1}:=\intop_{-1}^{1}\frac{\prod_{i=1}^{n}\mu_{i}d\mu_{i}}{\sqrt{1-\sum_{i=1}^{n}\mu_{i}^{2}}}.\label{integral1}
\end{equation}
Note that, the integer $n$ has not specified as being odd or even.
Considering the following parameterizations of the $\mu_{i}$' s
\begin{eqnarray}
 &  & \mu_{1}=\cos\theta_{1}\nonumber \\
 &  & \mu_{2}=\sin\theta_{1}\cos\theta_{2}\nonumber \\
 &  & .\nonumber \\
 &  & .\nonumber \\
 &  & .\nonumber \\
 &  & \mu_{n-1}=\sin\theta_{1}...\sin\theta_{n-2}\cos\theta_{n-1}\nonumber \\
 &  & \mu_{n}=\sin\theta_{1}...\sin\theta_{n-1}\cos\theta_{n},\label{parameterization}
\end{eqnarray}
one can calculate the integral easily. First, let us compute the denominator.
We have

\begin{equation}
1-\sum_{i=1}^{n}\mu_{i}^{2}=1-\cos^{2}\theta_{1}-\sin^{2}\theta_{1}\cos^{2}\theta_{2}-...-\sin^{2}\theta_{1}...\sin^{2}\theta_{n-2}\cos^{2}\theta_{n-1}-\sin^{2}\theta_{1}...\sin^{2}\theta_{n-1}\cos^{2}\theta_{n},
\end{equation}
which reduces to the following simple expression
\begin{equation}
1-\sum_{i=1}^{n}\mu_{i}^{2}=\sin^{2}\theta_{1}\sin^{2}\theta_{2}...\sin^{2}\theta_{n-1}\sin^{2}\theta_{n}.
\end{equation}
Then, we can express the denominator of the integrand as
\begin{equation}
\sqrt{1-\sum_{i=1}^{n}\mu_{i}^{2}}=\sin\theta_{1}\sin\theta_{2}...\sin\theta_{n-1}\sin\theta_{n}.\label{piece1}
\end{equation}
Now let us compute the numerator. The multiplication of the $\mu_{i}$
parameters ,$\mu_{1}\mu_{2}...\mu_{n}$, in terms of the $\theta_{i}$'
s yields

\begin{equation}
\prod_{i=1}^{n}\mu_{i}=\cos\theta_{1}\sin^{n-1}\theta_{1}\cos\theta_{2}\sin^{n-2}\theta_{2}...\cos\theta_{n-1}\sin\theta_{n-1}\cos\theta_{n}.\label{piece2}
\end{equation}
Also, we need to calculate the $d\mu_{1}d\mu_{2}...d\mu_{n}$ term.
Since 
\begin{equation}
\prod_{i=1}^{n}d\mu_{i}=\det\begin{vmatrix}\frac{\partial\mu_{1}}{\partial\theta_{1}} & \frac{\partial\mu_{1}}{\partial\theta_{2}} & . & . & . & \frac{\partial\mu_{1}}{\partial\theta_{n}}\\
\frac{\partial\mu_{2}}{\partial\theta_{1}} & \frac{\partial\mu_{2}}{\partial\theta_{2}} & . & . & . & \frac{\partial\mu_{2}}{\partial\theta_{n}}\\
. & . &  &  &  & .\\
. &  & . &  &  & .\\
. &  &  & . &  & .\\
\frac{\partial\mu_{n}}{\partial\theta_{1}} & \frac{\partial\mu_{n}}{\partial\theta_{2}} & . & . & . & \frac{\partial\mu_{n}}{\partial\theta_{n}}
\end{vmatrix}d\theta_{1}d\theta_{2}...d\theta_{n},
\end{equation}
one ends up with
\begin{equation}
\prod_{i=1}^{n}d\mu_{i}=\sin^{n}\theta_{1}\sin^{n-1}\theta_{2}...\sin^{2}\theta_{n-1}\sin\theta_{n}d\theta_{1}d\theta_{2}...d\theta_{n}.\label{piece3}
\end{equation}
Collecting the pieces (\ref{piece2}, \ref{piece3}), one arrives
at the numerator as
\begin{equation}
\prod_{i=1}^{n}\mu_{i}d\mu_{i}=\cos\theta_{1}\sin^{2n-1}\theta_{1}\cos\theta_{2}\sin^{2n-3}\theta_{2}...\cos\theta_{n-1}\sin^{3}\theta_{n-1}\cos\theta_{n}\sin\theta_{n}d\theta_{1}d\theta_{2}...d\theta_{n}.
\end{equation}
Then, the integrand can be expressed in terms of the new parameters.
We have
\begin{equation}
\frac{\prod_{i=1}^{n}\mu_{i}d\mu_{i}}{\sqrt{1-\sum_{i=1}^{n}\mu_{i}^{2}}}=\cos\theta_{1}\sin^{2n-2}\theta_{1}\cos\theta_{2}\sin^{2n-4}\theta_{2}...\cos\theta_{n-1}\sin^{2}\theta_{n-1}\cos\theta_{n}d\theta_{1}d\theta_{2}...d\theta_{n}.\label{combinedpieces}
\end{equation}
Now, let us consider the boundaries of the integral. We can write
\begin{equation}
I_{1}=\intop_{-1}^{1}\frac{\prod_{i=1}^{n}\mu_{i}d\mu_{i}}{\sqrt{1-\sum_{i=1}^{n}\mu_{i}^{2}}}=2\intop_{0}^{1}\frac{\prod_{i=1}^{n}\mu_{i}d\mu_{i}}{\sqrt{1-\sum_{i=1}^{n}\mu_{i}^{2}}},
\end{equation}
since the integrand is an even function of the $\mu_{i}$' s. The
integral then becomes
\begin{equation}
I_{1}=2\intop_{-\pi/2}^{0}...\intop_{-\pi/2}^{0}\cos\theta_{1}\sin^{2n-2}\theta_{1}\cos\theta_{2}\sin^{2n-4}\theta_{2}...\cos\theta_{n-1}\sin^{2}\theta_{n-1}\cos\theta_{n}d\theta_{1}d\theta_{2}...d\theta_{n}.
\end{equation}
So, one obtains the following equation
\begin{equation}
I_{1}=2\intop_{-\pi/2}^{0}\cos\theta_{1}\sin^{2n-2}\theta_{1}d\theta_{1}\intop_{-\pi/2}^{0}\cos\theta_{2}\sin^{2n-4}\theta_{2}d\theta_{2}...\intop_{-\pi/2}^{0}\cos\theta_{n-1}\sin^{2}\theta_{n-1}d\theta_{n-1}\intop_{-\pi/2}^{0}\cos\theta_{n}d\theta_{n},
\end{equation}
which is now in a familiar form. The $\theta_{n}$ integral yields
one and then one has
\begin{equation}
I_{1}=2\intop_{-\pi/2}^{0}\cos\theta_{1}\sin^{2n-2}\theta_{1}d\theta_{1}\intop_{-\pi/2}^{0}\cos\theta_{2}\sin^{2n-4}\theta_{2}d\theta_{2}...\intop_{-\pi/2}^{0}\cos\theta_{n-1}\sin^{2}\theta_{n-1}d\theta_{n-1}.
\end{equation}
Defining the $y_{i}$ functions $y_{i}:=\sin\theta_{i}$, where $i=1,2,...,n-1$,
we can rewrite the last equation as
\begin{equation}
I_{1}=2\intop_{-1}^{0}y_{1}^{2n-2}dy_{1}\intop_{-1}^{0}y_{2}^{2n-4}dy_{2}...\intop_{-1}^{0}y_{n-1}^{2}dy_{n-1}.
\end{equation}
Finally, we can express the integral given in (\ref{integral1}) as
\begin{equation}
I_{1}=2\frac{1}{(2n-1)(2n-3)...5.3}=\frac{2}{(2n-1)!!}.
\end{equation}
We also need to compute the following integral
\begin{equation}
I_{2}:=\intop_{-1}^{1}\frac{\prod_{i=1}^{n}\mu_{i}d\mu_{i}}{\sqrt{1-\sum_{i=1}^{n}\mu_{i}^{2}}}\sum_{j=1}^{n}\alpha_{j}\mu_{j}^{2},\label{integral2}
\end{equation}
where the $\frac{\prod_{i=1}^{n}\mu_{i}d\mu_{i}}{\sqrt{1-\sum_{i=1}^{n}\mu_{i}^{2}}}$
piece was given in equation (\ref{combinedpieces}). The remaining
piece reads

\begin{equation}
\sum_{j=1}^{n}\alpha_{j}\mu_{j}^{2}=\alpha_{1}\cos^{2}\theta_{1}+\alpha_{2}\sin^{2}\theta_{1}\cos^{2}\theta_{2}+...+\alpha_{n}\sin^{2}\theta_{1}...\sin^{2}\theta_{n-1}\cos^{2}\theta_{n}.
\end{equation}
Then, we can express the second integral (\ref{integral2}) as
\begin{eqnarray}
 &  & I_{2}=2\biggl(\alpha_{1}\intop_{-\pi/2}^{0}\cos^{3}\theta_{1}\sin^{2n-2}\theta_{1}d\theta_{1}\intop_{-\pi/2}^{0}\cos\theta_{2}\sin^{2n-4}\theta_{2}d\theta_{2}...\intop_{-\pi/2}^{0}\cos\theta_{n-1}\sin^{2}\theta_{n-1}d\theta_{n-1}\nonumber \\
 &  & ~~~~~~+\alpha_{2}\intop_{-\pi/2}^{0}\cos\theta_{1}\sin^{2n}\theta_{1}d\theta_{1}\intop_{-\pi/2}^{0}\cos^{3}\theta_{2}\sin^{2n-2}\theta_{2}d\theta_{2}...\intop_{-\pi/2}^{0}\cos\theta_{n-1}\sin^{2}\theta_{n-1}d\theta_{n-1}\nonumber \\
 &  & ~~~~~...+\alpha_{n}\intop_{-\pi/2}^{0}\cos\theta_{1}\sin^{2n}\theta_{1}d\theta_{1}...\intop_{-\pi/2}^{0}\cos\theta_{n-1}\sin^{4}\theta_{n-1}d\theta_{n-1}\intop_{-\pi/2}^{0}\cos^{3}\theta_{n}d\theta_{n}\biggr).\label{second integral}
\end{eqnarray}
For simplicity, let us focus on the first integral on the right hand
side of the last equation. Using the $y_{i}$ functions, $y_{1}=\sin\theta_{1},y_{2}=\sin\theta_{2}...y_{n-1}=\sin\theta_{n-1}$,
again this piece yields $\frac{2}{(2n+1)!!}$ . Similarly, the additional
pieces give the same value. So then, one finds 
\begin{equation}
I_{2}=2\left(\alpha_{1}\frac{2}{(2n+1)!!}+\alpha_{2}\frac{2}{(2n+1)!!}+...+\alpha_{n}\frac{2}{(2n+1)!!}\right)=\frac{4}{(2n+1)!!}\sum_{j=1}^{n}\alpha_{j}.
\end{equation}
Note that, due to the equality of the pieces, we can express the following
result
\begin{equation}
I_{3}:=\intop_{-1}^{1}\frac{\prod_{i=1}^{n}\mu_{i}d\mu_{i}}{\sqrt{1-\sum_{i=1}^{n}\mu_{i}^{2}}}\alpha_{j}\mu_{j}^{2}=\frac{4}{(2n+1)!!}\alpha_{j},\label{integral3}
\end{equation}
which also denotes any arbitrary piece of the second integral.

\section*{APPENDIX C: ENERGY AND ANGULAR MOMENTA IN $D$ DIMENSIONS}

In this section, we give the construction of the energy and angular
momentum of the for the even dimensional Kerr-AdS metric solutions
(\ref{eq:KS}) in cosmological Einstein theory. To compute the energy,
we need to use the energy Killing vector, $\bar{\xi}^{\mu}=(-1,\vec{0})$
and compute the integrand in charge expression (\ref{newcharges})
for the given Killing vector. We have
\begin{equation}
\left(R^{r0}\thinspace_{\beta\sigma}\right)^{\left(1\right)}\bar{\nabla}^{\beta}\bar{\xi}^{\sigma}=\left(\left(R^{r}\thinspace_{\rho\beta\sigma}\right)^{\left(1\right)}\overline{g}^{\rho0}-\overline{R}^{r}\thinspace_{\rho\beta\sigma}h^{\rho0}\right)\bar{\nabla}^{\beta}\bar{\xi}^{\sigma},
\end{equation}
where the linearized Riemann tensor reads
\begin{equation}
\left(R^{r}\thinspace_{\rho\beta\sigma}\right)^{\left(1\right)}=\overline{\nabla}_{\beta}(\Gamma_{\rho\sigma}^{r})^{(1)}-\overline{\nabla}_{\sigma}(\Gamma_{\rho\beta}^{r})^{(1)},
\end{equation}
and the first order expansion of the Christoffel symbol is
\begin{equation}
(\Gamma_{\rho\sigma}^{r})^{(1)}=\frac{1}{2}(\overline{\nabla}h_{\sigma}^{r}+\overline{\nabla}_{\sigma}h_{\rho}^{r}-\overline{\nabla}^{r}h_{\rho\sigma}).
\end{equation}
 Using the antisymmetry of the indices, $\beta$ and $\sigma$, one
obtains
\begin{equation}
\left(R^{r0}\thinspace_{\beta\sigma}\right)^{\left(1\right)}\bar{\nabla}^{\beta}\bar{\xi}^{\sigma}=2\overline{g}^{00}\bar{\nabla}^{\beta}\bar{\xi}^{\sigma}\overline{\nabla}_{\beta}(\Gamma_{\rho\sigma}^{r})^{(1)}-\overline{R}^{r}\thinspace_{\rho\beta\sigma}h^{\rho0}\bar{\nabla}^{\beta}\bar{\xi}^{\sigma},\label{abcd}
\end{equation}
where
\begin{equation}
\bar{\nabla}^{\beta}\bar{\xi}^{\sigma}=\frac{1}{2}\overline{g}^{\beta\nu}\overline{g}^{\sigma\gamma}(\partial_{\gamma}\overline{g}_{\nu0}-\partial_{\nu}\overline{g}_{\gamma0})
\end{equation}
for the Killing vector $\bar{\xi}^{\mu}=(-1,\vec{0})$. Then one has
\begin{equation}
\left(R^{r0}\thinspace_{\beta\sigma}\right)^{\left(1\right)}\bar{\nabla}^{\beta}\bar{\xi}^{\sigma}=(\overline{g}^{00})^{2}\overline{g}^{\beta\nu}\partial_{\nu}\overline{g}_{00}\left(\overline{\nabla}_{0}(\Gamma_{0\beta}^{r})^{(1)}-\overline{\nabla}_{\beta}(\Gamma_{00}^{r})^{(1)}\right)+\partial_{\nu}\overline{g}_{00}\overline{R}^{r}\thinspace_{\rho}\thinspace^{\nu0}h^{\rho0}.
\end{equation}
Expressing the linearized Christoffel connection in terms of the linear
order metric perturbations, one arrives at
\begin{align}
 & \left(R^{r0}\thinspace_{\beta\sigma}\right)^{\left(1\right)}\bar{\nabla}^{\beta}\bar{\xi}^{\sigma}=\frac{1}{2}\partial_{\nu}\overline{g}_{00}\left(\overline{R}^{0\nu0\rho}h_{\rho}^{r}-\overline{R}^{0\nu r\rho}h_{\rho}^{0}\right)\label{khgv-1}\\
 & ~~~~~~~~~~~~~~~~~~~~~+\frac{1}{2}(\overline{g}^{00})^{2}\overline{g}^{rr}\overline{g}^{\beta\nu}\partial_{\nu}\overline{g}_{00}\left(\overline{\nabla}_{0}\overline{\nabla}_{0}h_{\beta r}+\overline{\nabla}_{\beta}\overline{\nabla}_{r}h_{00}-\overline{\nabla}_{0}\overline{\nabla}_{r}h_{\beta0}-\overline{\nabla}_{\beta}\overline{\nabla}_{0}h_{0r}\right).\nonumber 
\end{align}
Since $\overline{g}_{00}$ is a function of $r$ and $\mu_{i}$' s,
we obtain the following 
\begin{eqnarray}
 &  & \left(R^{r0}\thinspace_{\beta\sigma}\right)^{\left(1\right)}\bar{\nabla}^{\beta}\bar{\xi}^{\sigma}=\frac{1}{2}\partial_{r}\overline{g}_{00}\left(\overline{R}^{0r0\rho}h_{\rho}^{r}-\overline{R}^{0rr\rho}h_{\rho}^{0}\right)+\frac{1}{2}\partial_{\mu_{i}}\overline{g}_{00}\left(\overline{R}^{0\mu_{i}0\rho}h_{\rho}^{r}-\overline{R}^{0\mu_{i}r\rho}h_{\rho}^{0}\right)\label{ggg-1}\\
 &  & ~~~~~~~~~~~~~~~~~~~~~~+\frac{1}{2}(\overline{g}^{00}\overline{g}^{rr})^{2}\partial_{r}\overline{g}_{00}\left(\overline{\nabla}_{0}\overline{\nabla}_{0}h_{rr}+\overline{\nabla}_{r}\overline{\nabla}_{r}h_{00}-\overline{\nabla}_{0}\overline{\nabla}_{r}h_{r0}-\overline{\nabla}_{r}\overline{\nabla}_{0}h_{0r}\right)\nonumber \\
 &  & ~~~~~~~~~~~~~~~~~~~~~~+\frac{1}{2}(\overline{g}^{00})^{2}\overline{g}^{rr}\overline{g}^{\mu_{i}\mu_{j}}\partial_{\mu_{i}}\overline{g}_{00}\left(\overline{\nabla}_{0}\overline{\nabla}_{0}h_{\mu_{j}r}+\overline{\nabla}_{\mu_{j}}\overline{\nabla}_{r}h_{00}-\overline{\nabla}_{0}\overline{\nabla}_{r}h_{\mu_{j}0}-\overline{\nabla}_{\mu_{j}}\overline{\nabla}_{0}h_{0r}\right).\nonumber 
\end{eqnarray}
We have $\overline{g}_{00}=W\Lambda r^{2}$ and $\overline{g}_{rr}=-\frac{1}{\Lambda r^{2}}$
when we take the $r\rightarrow\infty$ limit. Using the constraint
$\sum_{i=1}^{N+\epsilon}\mu_{i}^{2}=1$, one obtains $U=r^{D-3}$
and from the equations (\ref{k_mu's}, \ref{h_munu}) we can express
\begin{equation}
h_{00}=2MW^{2}r^{3-D},~~~~~~~~~~~h_{rr}=\frac{2M}{\Lambda^{2}}r^{-D-1}.\label{h_rrandh_00}
\end{equation}
Note that, from equation (\ref{k_mu's}) we have $k_{\mu_{i}}=0$,
which yields $h_{\nu\mu_{i}}=0$. Let us calculate the right hand
side of the equation (\ref{ggg-1}) term by term. The third and the
fourth terms vanish and the first two terms yield

\begin{equation}
\frac{1}{2}\partial_{r}\overline{g}_{00}\left(\overline{R}^{0r0\rho}h_{\rho}^{r}-\overline{R}^{0rr\rho}h_{\rho}^{0}\right)=2M\Lambda r^{2-D}(1-W).
\end{equation}
To obtain the second line, first express the covariant derivatives
in terms of the partial derivatives and the background Christoffel
symbol. The first piece yields
\begin{equation}
\overline{\nabla}_{0}\overline{\nabla}_{0}h_{rr}=-\overline{\Gamma}_{00}^{r}\partial_{r}h_{rr}+2\overline{\Gamma}_{00}^{r}\overline{\Gamma}_{rr}^{r}h_{rr}+2\overline{\Gamma}_{r0}^{0}\overline{\Gamma}_{00}^{r}h_{rr}+2\overline{\Gamma}_{r0}^{0}\overline{\Gamma}_{0r}^{0}h_{00}
\end{equation}
and the second one reads
\begin{equation}
\overline{\nabla}_{r}\overline{\nabla}_{r}h_{00}=\partial_{r}\partial_{r}h_{00}-2\partial_{r}(\overline{\Gamma}_{r0}^{0}h_{00})-\overline{\Gamma}_{rr}^{r}\partial_{r}h_{00}-2\overline{\Gamma}_{0r}^{0}\partial_{r}h_{00}+2\overline{\Gamma}_{r0}^{0}\overline{\Gamma}_{rr}^{r}h_{00}+4\overline{\Gamma}_{r0}^{0}\overline{\Gamma}_{0r}^{0}h_{00},
\end{equation}
also the third and the last pieces respectively yield
\begin{equation}
\overline{\nabla}_{0}\overline{\nabla}_{r}h_{0r}=\overline{\Gamma}_{r0}^{0}\overline{\Gamma}_{00}^{r}h_{rr}+3\overline{\Gamma}_{r0}^{0}\overline{\Gamma}_{0r}^{0}h_{00}-\overline{\Gamma}_{00}^{r}\partial_{r}h_{rr}+2\overline{\Gamma}_{00}^{r}\overline{\Gamma}_{rr}^{r}h_{rr}-\overline{\Gamma}_{0r}^{0}\partial_{r}h_{00},
\end{equation}
and
\begin{equation}
\overline{\nabla}_{r}\overline{\nabla}_{0}h_{r0}=2\overline{\Gamma}_{r0}^{0}\overline{\Gamma}_{00}^{r}h_{rr}+2\overline{\Gamma}_{r0}^{0}\overline{\Gamma}_{0r}^{0}h_{00}+\overline{\Gamma}_{00}^{r}\overline{\Gamma}_{rr}^{r}h_{rr}+\overline{\Gamma}_{r0}^{0}\overline{\Gamma}_{rr}^{r}h_{00}-\partial_{r}(\overline{\Gamma}_{00}^{r}h_{rr})-\partial_{r}(\overline{\Gamma}_{r0}^{0}h_{00}).
\end{equation}
Inserting the pieces, one arrives at 
\begin{eqnarray}
 &  & \overline{\nabla}_{0}\overline{\nabla}_{0}h_{rr}+\overline{\nabla}_{r}\overline{\nabla}_{r}h_{00}-\overline{\nabla}_{0}\overline{\nabla}_{r}h_{r0}-\overline{\nabla}_{r}\overline{\nabla}_{0}h_{0r}=\partial_{r}\partial_{r}h_{00}-\overline{\Gamma}_{rr}^{r}\partial_{r}h_{00}-2\overline{\Gamma}_{0r}^{0}\partial_{r}h_{00}\nonumber \\
 &  & +\overline{\Gamma}_{00}^{r}\partial_{r}h_{rr}+h_{rr}(\partial_{r}\overline{\Gamma}_{00}^{r}-\overline{\Gamma}_{rr}^{r}\overline{\Gamma}_{00}^{r}-\overline{\Gamma}_{0r}^{0}\overline{\Gamma}_{00}^{r})+h_{00}(-\partial_{r}\overline{\Gamma}_{r0}^{0}+\overline{\Gamma}_{rr}^{r}\overline{\Gamma}_{r0}^{0}+\overline{\Gamma}_{0r}^{0}\overline{\Gamma}_{r0}^{0}),\label{second piece-2}
\end{eqnarray}
where
\begin{equation}
\overline{\varGamma}_{\nu\rho}^{\mu}=\frac{1}{2}\overline{g}^{\mu\sigma}(\partial_{\nu}\overline{g}_{\rho\sigma}+\partial_{\rho}\overline{g}_{\nu\sigma}-\partial_{\sigma}\overline{g}_{\rho\nu}).\label{backgroundchristoffel}
\end{equation}
After a straightforward calculation, one ends up with 
\begin{equation}
\overline{\nabla}_{0}\overline{\nabla}_{0}h_{rr}+\overline{\nabla}_{r}\overline{\nabla}_{r}h_{00}-\overline{\nabla}_{0}\overline{\nabla}_{r}h_{r0}-\overline{\nabla}_{r}\overline{\nabla}_{0}h_{0r}=2MWr^{1-D}(WD^{2}-4WD+4W+2-D).
\end{equation}
So then, the second line in equation (\ref{ggg-1}) becomes 
\begin{eqnarray}
 &  & \frac{1}{2}(\overline{g}^{00}\overline{g}^{rr})^{2}\partial_{r}\overline{g}_{00}\left(\overline{\nabla}_{0}\overline{\nabla}_{0}h_{rr}+\overline{\nabla}_{r}\overline{\nabla}_{r}h_{00}-\overline{\nabla}_{0}\overline{\nabla}_{r}h_{r0}-\overline{\nabla}_{r}\overline{\nabla}_{0}h_{0r}\right)\label{secondline-1}\\
 &  & ~~~~~~~~~~~~~~~~~~~~~~~~~~~~~~~=2M\Lambda r^{2-D}(WD^{2}-4WD+4W+2-D).\nonumber 
\end{eqnarray}
Now, let us compute the last line. The first two pieces read
\begin{equation}
\overline{\nabla}_{0}\overline{\nabla}_{0}h_{\mu_{j}r}=\overline{\Gamma}_{00}^{\mu_{k}}\overline{\Gamma}_{\mu_{k}\mu_{j}}^{r}h_{rr}+\overline{\Gamma}_{0\mu_{j}}^{0}\overline{\Gamma}_{00}^{r}h_{rr}+2\overline{\Gamma}_{r0}^{0}\overline{\Gamma}_{0\mu_{j}}^{0}h_{00},
\end{equation}
and 
\begin{equation}
\overline{\nabla}_{\mu_{j}}\overline{\nabla}_{r}h_{00}=\partial_{\mu_{j}}\partial_{r}h_{00}-2\partial_{\mu_{j}}(\overline{\Gamma}_{r0}^{0}h_{00})-\overline{\Gamma}_{r\mu_{j}}^{\mu_{k}}\partial_{\mathbf{\mu_{k}}}h_{00}+2\overline{\Gamma}_{r\mu_{j}}^{\mu_{k}}\overline{\Gamma}_{0\mu_{k}}^{0}h_{00}-2\overline{\Gamma}_{0\mu_{j}}^{0}\partial_{r}h_{00}+4\overline{\Gamma}_{0\mu_{j}}^{0}\overline{\Gamma}_{0r}^{0}h_{00}.
\end{equation}
The third and the last terms yield
\begin{equation}
\overline{\nabla}_{0}\overline{\nabla}_{r}h_{0\mu_{j}}=3\overline{\Gamma}_{r0}^{0}\overline{\Gamma}_{0\mu_{j}}^{0}h_{00}-\overline{\Gamma}_{0\mu_{j}}^{0}\partial_{r}h_{00},
\end{equation}
and 
\begin{equation}
\overline{\nabla}_{\mu_{j}}\overline{\nabla}_{0}h_{0r}=-\partial_{\mu_{j}}(\overline{\Gamma}_{00}^{r}h_{rr})-\partial_{\mu_{j}}(\overline{\Gamma}_{r0}^{0}h_{00})+2\overline{\Gamma}_{0\mu_{j}}^{0}\overline{\Gamma}_{00}^{r}h_{rr}+2\overline{\Gamma}_{r0}^{0}\overline{\Gamma}_{0\mu_{j}}^{0}h_{00}+\overline{\Gamma}_{r\mu_{j}}^{\mu_{k}}\overline{\Gamma}_{0\mu_{k}}^{0}h_{00}.
\end{equation}
Collecting the pieces, we end up with
\begin{eqnarray}
 &  & \overline{\nabla}_{0}\overline{\nabla}_{0}h_{\mu_{j}r}+\overline{\nabla}_{\mu_{j}}\overline{\nabla}_{r}h_{00}-\overline{\nabla}_{0}\overline{\nabla}_{r}h_{\mu_{j}0}-\overline{\nabla}_{\mu_{j}}\overline{\nabla}_{0}h_{0r}\\
 &  & =\overline{\Gamma}_{00}^{\mu_{k}}\overline{\Gamma}_{\mu_{k}\mu_{j}}^{r}h_{rr}-\overline{\Gamma}_{0\mu_{j}}^{0}\overline{\Gamma}_{00}^{r}h_{rr}+\partial_{\mu_{j}}\partial_{r}h_{00}-h_{00}\partial_{\mu_{j}}\overline{\Gamma}_{r0}^{0}-\overline{\Gamma}_{r0}^{0}\partial_{\mu_{j}}h_{00}-\overline{\Gamma}_{r\mu_{j}}^{\mu_{k}}\partial_{\mathbf{\mu_{k}}}h_{00}\nonumber \\
 &  & +\overline{\Gamma}_{r\mu_{j}}^{\mu_{k}}\overline{\Gamma}_{0\mu_{k}}^{0}h_{00}-\overline{\Gamma}_{0\mu_{j}}^{0}\partial_{r}h_{00}+\overline{\Gamma}_{0\mu_{j}}^{0}\overline{\Gamma}_{0r}^{0}h_{00}+h_{rr}\partial_{\mu_{j}}\overline{\Gamma}_{00}^{r}.\nonumber 
\end{eqnarray}
Using the corresponding components of the linear order metric perturbation
(\ref{h_rrandh_00}), $\partial_{r}\overline{g}_{\mu_{i}\mu_{j}}=\frac{2}{r}\overline{g}_{\mu_{i}\mu_{j}}$
and $\partial_{\mu_{i}}\overline{g}_{00}=\Lambda r^{2}\partial_{\mu_{i}}W$
one obtains
\begin{equation}
\overline{\nabla}_{0}\overline{\nabla}_{0}h_{\mu_{j}r}+\overline{\nabla}_{\mu_{j}}\overline{\nabla}_{r}h_{00}-\overline{\nabla}_{0}\overline{\nabla}_{r}h_{\mu_{j}0}-\overline{\nabla}_{\mu_{j}}\overline{\nabla}_{0}h_{0r}=3MWr^{2-D}\partial_{\mu_{j}}W(1-D).
\end{equation}
Then, we can express the last line in (\ref{ggg-1}) as 
\begin{eqnarray}
 &  & \frac{1}{2}(\overline{g}^{00})^{2}\overline{g}^{rr}\overline{g}^{\mu_{i}\mu_{j}}\partial_{\mu_{i}}\overline{g}_{00}\left(\overline{\nabla}_{0}\overline{\nabla}_{0}h_{\mu_{j}r}+\overline{\nabla}_{\mu_{j}}\overline{\nabla}_{r}h_{00}-\overline{\nabla}_{0}\overline{\nabla}_{r}h_{\mu_{j}0}-\overline{\nabla}_{\mu_{j}}\overline{\nabla}_{0}h_{0r}\right)\nonumber \\
 &  & ~~~~~~~~~~~~~~~~~~~~~~~~~~~~~~~~~~~~~~~=-\frac{3M}{2W}r^{2-D}(1-D)\overline{g}^{\mu_{i}\mu_{j}}\partial_{\mu_{i}}W\partial_{\mu_{j}}W.
\end{eqnarray}
We need to compute the $\overline{g}^{\mu_{i}\mu_{j}}\partial_{\mu_{i}}W\partial_{\mu_{j}}W$
term, which is complicated due to the inverse metric. In order to
simplify the calculation, we can use the explicit form of the Riemann
tensor. Let us consider the $\overline{R}^{0}\thinspace_{\phi_{i}0\phi_{j}}$
component. We can express
\begin{equation}
\overline{R}^{0}\thinspace_{\phi_{i}0\phi_{j}}=\Lambda\overline{g}_{\phi_{i}\phi_{j}}=\overline{\Gamma}_{0r}^{0}\overline{\Gamma}_{\phi_{i}\phi_{j}}^{r}+\overline{\Gamma}_{0\mu_{k}}^{0}\overline{\Gamma}_{\phi_{i}\phi_{j}}^{\mu_{k}},
\end{equation}
where the first equality comes from the equation (\ref{dsriemann-ricci}).
Using $\overline{\Gamma}_{0r}^{0}\overline{\Gamma}_{\phi_{i}\phi_{j}}^{r}=\Lambda\overline{g}_{\phi_{i}\phi_{j}}$,
we arrive at 
\begin{equation}
\overline{\Gamma}_{0\mu_{k}}^{0}\overline{\Gamma}_{\phi_{i}\phi_{j}}^{\mu_{k}}=0.
\end{equation}
In terms of the components of the background metric tensor, the last
equation can be written as 
\begin{equation}
\overline{\Gamma}_{0\mu_{k}}^{0}\overline{\Gamma}_{\phi_{i}\phi_{j}}^{\mu_{k}}=-\frac{1}{4W}\partial_{k}W\overline{g}^{\mu_{k}\mu_{l}}\partial_{\mu_{l}}\overline{g}_{\phi_{i}\phi_{j}}=0,
\end{equation}
which yields the following identity
\begin{equation}
\overline{g}^{\mu_{k}\mu_{l}}\partial_{\mu_{l}}\overline{g}_{\phi_{i}\phi_{j}}\partial_{\mu_{k}}W=0.\label{identity-1}
\end{equation}
The $r\rightarrow\infty$ limit of the $\overline{g}_{\phi_{i}\phi_{j}}$
components reads $\frac{r^{2}\mu_{i}^{2}}{1+\Lambda a_{i}^{2}}\delta_{ij}$,
where $i=1,...N$. Then one finds, $\overline{g}^{\mu_{i}\mu_{j}}\partial_{\mu_{i}}W\partial_{\mu_{j}}W=0$.
So, the third line in equation (\ref{ggg-1}) has no contribution
to the energy. Adding the first two terms and the second line, the
non-vanishing terms, we arrive at
\begin{equation}
\left(R^{r0}\thinspace_{\beta\sigma}\right)^{\left(1\right)}\bar{\nabla}^{\beta}\bar{\xi}^{\sigma}=2M\Lambda r^{2-D}(D-3)\Bigl(W(D-1)-1\Bigr).
\end{equation}
Substituting this result in (\ref{chargeformulaforkerrschildform}),
one can express the energy as 
\begin{equation}
Q=\frac{M}{2G_{D}\Omega_{D-2}}\int_{\partial\bar{\Sigma}}dS_{r}\,r^{2-D}\Bigl(W(D-1)-1\Bigr),
\end{equation}
where one can write
\begin{equation}
W(D-1)-1=D-2-(D-1)\sum_{i=1}^{N}\frac{\Lambda\mu_{i}^{2}a_{i}^{2}}{1+\Lambda a_{i}^{2}}
\end{equation}
and in the even dimensional case the solid angle reads
\begin{equation}
\Omega_{D-2}=\frac{2^{\frac{D}{2}}\pi^{\frac{D}{2}-1}}{(D-3)!!}.\label{solid angle even}
\end{equation}
Taking $G_{D}=1$ and inserting the $\sqrt{-\overline{g}}$ , it was
given in equation (\ref{determinantlast}) in Appendix A, one has
\begin{equation}
Q=\frac{M(D-3)!!}{2^{D+1}\pi^{\frac{D}{2}-1}}\int_{-1}^{1}\prod_{i=1}^{N}\frac{\mu_{i}d\mu_{i}}{(1+\Lambda a_{i}^{2})\sqrt{1-\sum_{i=1}^{N}\mu_{i}^{2}}}\Bigl(D-2-(D-1)\sum_{j=1}^{N}\frac{\Lambda\mu_{j}^{2}a_{j}^{2}}{1+\Lambda a_{j}^{2}}\Bigr)\int_{0}^{2\pi}\prod_{k=1}^{N}d\phi_{k},
\end{equation}
where the $\phi_{k}$ integrals yield $(2\pi)^{N}=(2\pi)^{\frac{D}{2}-1}$.
Defining the function $\Xi$ 
\begin{equation}
\Xi\equiv\prod_{i=1}^{N}\left(1+\Lambda a_{i}^{2}\right),\label{=00005CXi-1}
\end{equation}
we can express the energy corresponds to the Killing vector $\overline{\xi}^{\mu}=(-1,\overrightarrow{0})$
as
\begin{equation}
Q=\frac{M(D-3)!!}{4\Xi}\int_{-1}^{1}\frac{\prod_{i=1}^{N}\mu_{i}d\mu_{i}}{\sqrt{1-\sum_{i=1}^{N}\mu_{i}^{2}}}\Bigl(D-2-(D-1)\sum_{j=1}^{N}\frac{\Lambda\mu_{j}^{2}a_{j}^{2}}{1+\Lambda a_{j}^{2}}\Bigr).
\end{equation}
To take the integral, use the equations (\ref{integral1}, \ref{integral2})
given in Appendix B. We can write
\begin{equation}
\int_{-1}^{1}\frac{\prod_{i=1}^{N}\mu_{i}d\mu_{i}}{\sqrt{1-\sum_{i=1}^{N}\mu_{i}^{2}}}=\frac{2}{(2N-1)!!}=\frac{2}{(D-3)!!}
\end{equation}
and also
\begin{equation}
\int_{-1}^{1}\frac{\prod_{i=1}^{N}\mu_{i}d\mu_{i}}{\sqrt{1-\sum_{i=1}^{N}\mu_{i}^{2}}}\sum_{j=1}^{N}\frac{\Lambda\mu_{j}^{2}a_{j}^{2}}{1+\Lambda a_{j}^{2}}=\frac{4}{(2N+1)!!}\sum_{j=1}^{N}\frac{\Lambda a_{j}^{2}}{1+\Lambda a_{j}^{2}}=\frac{4}{(D-1)!!}\sum_{j=1}^{N}\frac{\Lambda a_{j}^{2}}{1+\Lambda a_{j}^{2}}.
\end{equation}
Then, the energy can be written as 
\begin{equation}
E=\frac{M}{\Xi}\Bigl(\frac{D-2}{2}-\sum_{j=1}^{N}\frac{\Lambda a_{j}^{2}}{1+\Lambda a_{j}^{2}}\Bigr),
\end{equation}
where we can express
\begin{equation}
\frac{D-2}{2}=\sum_{j=1}^{N}\frac{1+\Lambda a_{j}^{2}}{1+\Lambda a_{j}^{2}}.
\end{equation}
Finally, for the even dimensional case one ends up with 
\begin{equation}
E=\frac{M}{\Xi}\sum_{i=1}^{N}\frac{1}{\Xi_{i}},
\end{equation}
where we have defined the $\Xi_{i}$ as follows
\begin{equation}
\Xi_{i}\equiv1+\Lambda a_{i}^{2}.
\end{equation}
Similarly, one can compute the energy of the odd dimensional Kerr-Ads
black holes. Combining the results, we arrive at the energy of the
$D$-dimensional rotating black hole as
\begin{equation}
E=\frac{M}{\Xi}\sum_{i=1}^{N}\left(\frac{1}{\Xi_{i}}-\frac{1}{2}\left(1-\epsilon\right)\right).\label{eq:finalen-1-1}
\end{equation}
 To compute the angular momentum, one needs to perform a similar computation.
This time, we consider the Killing vector $\xi_{(i)}^{\mu}=\left(0,...,0,1_{i},0,...,0\right)$,
where the only non-zero term is the $i^{\mbox{th}}$ $\phi$ component.
One has
\begin{equation}
\overline{\nabla}^{\beta}\overline{\xi}^{\sigma}=\frac{1}{2}(\overline{g}^{\beta\nu}\overline{g}^{\sigma\phi_{j}}-\overline{g}^{\beta\phi_{j}}\overline{g}^{\sigma\nu})\partial_{\nu}\overline{g}_{\phi_{i}\phi_{j}},
\end{equation}
where only $r$ and $\mu_{i}$ derivatives of the $\overline{g}_{\phi_{i}\phi_{j}}$
component survive. Then we obtain
\begin{eqnarray}
 &  & \left(R^{r0}\thinspace_{\beta\sigma}\right)^{\left(1\right)}\bar{\nabla}^{\beta}\bar{\xi}^{\sigma}=\frac{1}{2}\partial_{r}\overline{g}_{\phi_{i}\phi_{j}}\left(\overline{R}^{r\phi_{j}0\rho}h_{\rho}^{r}-\overline{R}^{r\phi_{j}r\rho}h_{\rho}^{0}\right)+\frac{1}{2}\partial_{\mu_{i}}\overline{g}_{\phi_{i}\phi_{j}}\left(\overline{R}^{\mu_{i}\phi_{j}0\rho}h_{\rho}^{r}-\overline{R}^{\mu_{i}\phi_{j}r\rho}h_{\rho}^{0}\right)\nonumber \\
 &  & ~~~~~~~~~~~+\frac{1}{2}(\overline{g}^{rr})^{2}\overline{g}^{00}\overline{g}^{\phi_{j}\phi_{k}}\partial_{r}\overline{g}_{\phi_{i}\phi_{j}}\left(\overline{\nabla}_{r}\overline{\nabla}_{0}h_{\phi_{k}r}+\overline{\nabla}_{\phi_{k}}\overline{\nabla}_{r}h_{r0}-\overline{\nabla}_{r}\overline{\nabla}_{r}h_{\phi_{k}0}-\overline{\nabla}_{\phi_{k}}\overline{\nabla}_{0}h_{rr}\right)\label{gggg-1}\\
 &  & ~~~~~~~~~~~+\frac{1}{2}\overline{g}^{00}\overline{g}^{rr}\overline{g}^{\phi_{j}\phi_{k}}\overline{g}^{\mu_{l}\mu_{m}}\partial_{\mu_{m}}\overline{g}_{\phi_{i}\phi_{j}}\left(\overline{\nabla}_{\mu_{l}}\overline{\nabla}_{0}h_{\phi_{k}r}+\overline{\nabla}_{\phi_{k}}\overline{\nabla}_{r}h_{0\mu_{l}}-\overline{\nabla}_{\mu_{l}}\overline{\nabla}_{r}h_{\phi_{k}0}-\overline{\nabla}_{\phi_{k}}\overline{\nabla}_{0}h_{\mu_{l}r}\right).\nonumber 
\end{eqnarray}
The first four terms on the right hand side of the last equation reads
\begin{equation}
\frac{1}{2}\partial_{r}\overline{g}_{\phi_{i}\phi_{j}}\left(\overline{R}^{r\phi_{j}0\rho}h_{\rho}^{r}-\overline{R}^{r\phi_{j}r\rho}h_{\rho}^{0}\right)+\frac{1}{2}\partial_{\mu_{i}}\overline{g}_{\phi_{i}\phi_{j}}\left(\overline{R}^{\mu_{i}\phi_{j}0\rho}h_{\rho}^{r}-\overline{R}^{\mu_{i}\phi_{j}r\rho}h_{\rho}^{0}\right)=2M\Lambda r^{2-D}k_{\phi_{i}}.
\end{equation}
Let us compute the remaining terms. The first piece in the second
line yields
\begin{equation}
\overline{\nabla}_{r}\overline{\nabla}_{0}h_{\phi_{k}r}=-\partial_{r}(\overline{\Gamma}_{r0}^{0}h_{\phi_{k}0})+\overline{\Gamma}_{0r}^{0}\overline{\Gamma}_{0r}^{0}h_{0\phi_{k}}+\overline{\Gamma}_{r\phi_{k}}^{\phi_{i}}\overline{\Gamma}_{0r}^{0}h_{\phi_{i}0}+\overline{\Gamma}_{rr}^{r}\overline{\Gamma}_{0r}^{0}h_{\phi_{k}0},
\end{equation}
and the second one gives
\begin{equation}
\overline{\nabla}_{\phi_{k}}\overline{\nabla}_{r}h_{r0}=-\overline{\Gamma}_{r\phi_{k}}^{\phi_{i}}\partial_{r}h_{\phi_{i}0}+2\overline{\Gamma}_{r\phi_{k}}^{\phi_{i}}\overline{\Gamma}_{\phi_{i}r}^{\phi_{j}}h_{\phi_{j}0}+\overline{\Gamma}_{r\phi_{k}}^{\phi_{i}}\overline{\Gamma}_{0r}^{0}h_{\phi_{i}0}.
\end{equation}
The third and the fourth terms respectively read
\begin{eqnarray}
 &  & \overline{\nabla}_{r}\overline{\nabla}_{r}h_{\phi_{k}0}=\partial_{r}\partial_{r}h_{\phi_{k}0}-\partial_{r}(\overline{\Gamma}_{r0}^{0}h_{\phi_{k}0})-\partial_{r}(\overline{\Gamma}_{r\phi_{k}}^{\phi_{i}}h_{\phi_{i}0})-\overline{\Gamma}_{rr}^{r}\partial_{r}h_{\phi_{k}0}-\overline{\Gamma}_{0r}^{0}\partial_{r}h_{\phi_{k}0}-\overline{\Gamma}_{r\phi_{k}}^{\phi_{i}}\partial_{r}h_{\phi_{i}0}\nonumber \\
 &  & ~~~~~~~~~~~~~~+\overline{\Gamma}_{rr}^{r}\overline{\Gamma}_{0r}^{0}h_{\phi_{k}0}+\overline{\Gamma}_{r\phi_{k}}^{\phi_{i}}\overline{\Gamma}_{rr}^{r}h_{\phi_{i}0}+\overline{\Gamma}_{0r}^{0}\overline{\Gamma}_{0r}^{0}h_{0\phi_{k}}+2\overline{\Gamma}_{r\phi_{k}}^{\phi_{i}}\overline{\Gamma}_{0r}^{0}h_{\phi_{i}0}+\overline{\Gamma}_{r\phi_{k}}^{\phi_{i}}\overline{\Gamma}_{\phi_{i}r}^{\phi_{j}}h_{\phi_{j}0},\label{thethirdpiece-1}
\end{eqnarray}
and 
\begin{equation}
\overline{\nabla}_{\phi_{k}}\overline{\nabla}_{0}h_{rr}=2\overline{\Gamma}_{r\phi_{k}}^{\phi_{i}}\overline{\Gamma}_{0r}^{0}h_{\phi_{i}0}.
\end{equation}
Substituting the pieces, we obtain
\begin{eqnarray}
 &  & \overline{\nabla}_{r}\overline{\nabla}_{0}h_{\phi_{k}r}+\overline{\nabla}_{\phi_{k}}\overline{\nabla}_{r}h_{r0}-\overline{\nabla}_{r}\overline{\nabla}_{r}h_{\phi_{k}0}-\overline{\nabla}_{\phi_{k}}\overline{\nabla}_{0}h_{rr}=\overline{\Gamma}_{r\phi_{k}}^{\phi_{i}}\overline{\Gamma}_{\phi_{i}r}^{\phi_{j}}h_{\phi_{j}0}-\partial_{r}\partial_{r}h_{\phi_{k}0}\label{collected-2}\\
 &  & +h_{\phi_{i}0}\partial_{r}\overline{\Gamma}_{r\phi_{k}}^{\phi_{i}}+\overline{\Gamma}_{r\phi_{k}}^{\phi_{i}}\partial_{r}h_{\phi_{i}0}+\overline{\Gamma}_{rr}^{r}\partial_{r}h_{\phi_{k}0}+\overline{\Gamma}_{0r}^{0}\partial_{r}h_{\phi_{k}0}-\overline{\Gamma}_{r\phi_{k}}^{\phi_{i}}\overline{\Gamma}_{rr}^{r}h_{\phi_{i}0}-\overline{\Gamma}_{r\phi_{k}}^{\phi_{i}}\overline{\Gamma}_{0r}^{0}h_{\phi_{i}0}.\nonumber 
\end{eqnarray}
From the equations (\ref{h_munu}) and (\ref{k_mu's}) one has
\begin{equation}
h_{\phi_{i}0}=2MWr^{3-D}k_{\phi_{i}}.\label{h_0i-1}
\end{equation}
Using $\partial_{r}\overline{g}_{\phi_{i}\phi_{j}}=\frac{2}{r}\overline{g}_{\phi_{i}\phi_{j}}$
and taking the $r$ and $\mu_{i}$ derivatives, one arrives at
\begin{equation}
\overline{\nabla}_{r}\overline{\nabla}_{0}h_{\phi_{k}r}+\overline{\nabla}_{\phi_{k}}\overline{\nabla}_{r}h_{r0}-\overline{\nabla}_{r}\overline{\nabla}_{r}h_{\phi_{k}0}-\overline{\nabla}_{\phi_{k}}\overline{\nabla}_{0}h_{rr}=2MWr^{1-D}k_{k}(-D^{2}+4D-4).
\end{equation}
 Since one has $\frac{}{}(\overline{g}^{rr})^{2}\overline{g}^{00}\overline{g}^{\phi_{j}\phi_{k}}\partial_{r}\overline{g}_{\phi_{i}\phi_{j}}=2\Lambda W^{-1}r\delta_{i}^{k}$,
the second line in (\ref{gggg-1}) can be written as 
\begin{eqnarray}
 &  & \frac{1}{2}(\overline{g}^{rr})^{2}\overline{g}^{00}\overline{g}^{\phi_{j}\phi_{k}}\partial_{r}\overline{g}_{\phi_{i}\phi_{j}}\left(\overline{\nabla}_{r}\overline{\nabla}_{0}h_{\phi_{k}r}+\overline{\nabla}_{\phi_{k}}\overline{\nabla}_{r}h_{r0}-\overline{\nabla}_{r}\overline{\nabla}_{r}h_{\phi_{k}0}-\overline{\nabla}_{\phi_{k}}\overline{\nabla}_{0}h_{rr}\right)\nonumber \\
 &  & ~~~~~~~~~~~~~~~~~~~~~~~~~~~~~~~~~~~~~~~~~~~~~~~=2M\Lambda r^{2-D}k_{i}(-D^{2}+4D-4).\label{secondlineofangularmomentum-1}
\end{eqnarray}
Now, let us compute the last line in (\ref{gggg-1}). We have
\begin{equation}
\overline{\nabla}_{\mu_{l}}\overline{\nabla}_{0}h_{\phi_{k}r}=-\partial_{\mu_{l}}(\overline{\Gamma}_{r0}^{0}h_{\phi_{k}0})+\overline{\Gamma}_{0\mu_{l}}^{0}\overline{\Gamma}_{0r}^{0}h_{0\phi_{k}}+\overline{\Gamma}_{\mu_{l}\phi_{k}}^{\phi_{i}}\overline{\Gamma}_{0r}^{0}h_{\phi_{i}0}+\overline{\Gamma}_{\mu_{l}r}^{\mu_{j}}\overline{\Gamma}_{0\mu_{j}}^{0}h_{\phi_{k}0},
\end{equation}
and
\begin{equation}
\overline{\nabla}_{\phi_{k}}\overline{\nabla}_{r}h_{0\mu_{l}}=\overline{\Gamma}_{r\phi_{k}}^{\phi_{i}}\overline{\Gamma}_{\phi_{i}\mu_{l}}^{\phi_{j}}h_{\phi_{j}0}-\overline{\Gamma}_{\phi_{k}\mu_{l}}^{\phi_{i}}\partial_{r}h_{0\phi_{i}}+\overline{\Gamma}_{\mu_{l}\phi_{k}}^{\phi_{i}}\overline{\Gamma}_{0r}^{0}h_{\phi_{i}0}+\overline{\Gamma}_{\mu_{l}\phi_{k}}^{\phi_{i}}\overline{\Gamma}_{\phi_{i}r}^{\phi_{j}}h_{\phi_{j}0},
\end{equation}
and also
\begin{eqnarray}
 &  & \overline{\nabla}_{\mu_{m-l}}\overline{\nabla}_{r}h_{\phi_{k}0}=\partial_{\mu_{l}}\partial_{r}h_{\phi_{k}0}-\partial_{\mu_{l}}(\overline{\Gamma}_{r0}^{0}h_{\phi_{k}0})-\partial_{\mu_{l}}(\overline{\Gamma}_{r\phi_{k}}^{\phi_{i}}h_{\phi_{i}0})-\overline{\Gamma}_{\mu_{l}r}^{\mu_{i}}\partial_{\mu_{i}}h_{\phi_{k}0}\\
 &  & ~~~~~~~~~~~~~~~~~~+\overline{\Gamma}_{\mu_{l}r}^{\mu_{i}}\overline{\Gamma}_{0\mu_{i}}^{0}h_{\phi_{k}0}+\overline{\Gamma}_{\mu_{l}r}^{\mu_{i}}\overline{\Gamma}_{\phi_{k}\mu_{i}}^{\phi_{i}}h_{\phi_{i}0}-\overline{\Gamma}_{\mu_{l}0}^{0}\partial_{r}h_{\phi_{k}0}+\overline{\Gamma}_{\mu_{l}0}^{0}\overline{\Gamma}_{r0}^{0}h_{\phi_{k}0}\nonumber \\
 &  & ~~~~~~~~~~~~~~~~~~+\overline{\Gamma}_{\mu_{l}0}^{0}\overline{\Gamma}_{\phi_{k}r}^{\phi_{i}}h_{\phi_{i}0}-\overline{\Gamma}_{\phi_{k}\mu_{l}}^{\phi_{i}}\partial_{r}h_{0\phi_{i}}+\overline{\Gamma}_{\mu_{l}\phi_{k}}^{\phi_{i}}\overline{\Gamma}_{0r}^{0}h_{\phi_{i}0}+\overline{\Gamma}_{\mu_{l}\phi_{k}}^{\phi_{i}}\overline{\Gamma}_{\phi_{i}r}^{\phi_{j}}h_{\phi_{j}0},\nonumber 
\end{eqnarray}
and 
\begin{equation}
\overline{\nabla}_{\phi_{k}}\overline{\nabla}_{0}h_{r\mu_{l}}=\overline{\Gamma}_{\mu_{l}\phi_{k}}^{\phi_{i}}\overline{\Gamma}_{0r}^{0}h_{\phi_{i}0}+\overline{\Gamma}_{r\phi_{k}}^{\phi_{i}}\overline{\Gamma}_{\mu_{l}0}^{0}h_{\phi_{i}0}.
\end{equation}
Collecting the results, we obtain
\begin{eqnarray}
 &  & \overline{\nabla}_{\mu_{l}}\overline{\nabla}_{0}h_{\phi_{k}r}+\overline{\nabla}_{\phi_{k}}\overline{\nabla}_{r}h_{0\mu_{l}}-\overline{\nabla}_{\mu_{l}}\overline{\nabla}_{r}h_{\phi_{k}0}-\overline{\nabla}_{\phi_{k}}\overline{\nabla}_{0}h_{\mu_{l}r}=\overline{\Gamma}_{r\phi_{k}}^{\phi_{i}}\overline{\Gamma}_{\phi_{i}\mu_{l}}^{\phi_{j}}h_{\phi_{j}0}-\partial_{\mu_{l}}\partial_{r}h_{\phi_{k}0}\\
 &  & +\overline{\Gamma}_{r\phi_{k}}^{\phi_{i}}\partial_{\mu_{l}}h_{\phi_{i}0}+h_{\phi_{i}0}\partial_{\mu_{l}}\overline{\Gamma}_{r\phi_{k}}^{\phi_{i}}+\overline{\Gamma}_{\mu_{l}r}^{\mu_{i}}\partial_{\mu_{i}}h_{\phi_{k}0}-\overline{\Gamma}_{\mu_{l}r}^{\mu_{i}}\overline{\Gamma}_{\phi_{k}\mu_{i}}^{\phi_{i}}h_{\phi_{i}0}+\overline{\Gamma}_{\mu_{l}0}^{0}\partial_{r}h_{\phi_{k}0}-2\overline{\Gamma}_{\mu_{l}0}^{0}\overline{\Gamma}_{\phi_{k}r}^{\phi_{i}}h_{\phi_{i}0},\nonumber 
\end{eqnarray}
which yields
\begin{equation}
\overline{\nabla}_{\mu_{l}}\overline{\nabla}_{0}h_{\phi_{k}r}+\overline{\nabla}_{\phi_{k}}\overline{\nabla}_{r}h_{0\mu_{l}}-\overline{\nabla}_{\mu_{l}}\overline{\nabla}_{r}h_{\phi_{k}0}-\overline{\nabla}_{\phi_{k}}\overline{\nabla}_{0}h_{\mu_{l}r}=Mr^{2-D}(D-1)(2W\partial_{\mu_{l}}k_{\phi_{k}}+k_{\phi_{k}}\partial_{\mu_{l}}W)
\end{equation}
using (\ref{h_0i-1}). To compute the last line, we use 
\begin{equation}
\frac{1}{2}\overline{g}^{00}\overline{g}^{rr}\overline{g}^{\phi_{j}\phi_{k}}\overline{g}^{\mu_{l}\mu_{m}}\partial_{\mu_{m}}\overline{g}_{\phi_{i}\phi_{j}}=-\frac{1}{2W}\overline{g}^{\phi_{j}\phi_{k}}\overline{g}^{\mu_{l}\mu_{m}}\partial_{\mu_{m}}\overline{g}_{\phi_{i}\phi_{j}},
\end{equation}
and then, we can express 
\begin{eqnarray}
 &  & \frac{1}{2}\overline{g}^{00}\overline{g}^{rr}\overline{g}^{\phi_{j}\phi_{k}}\overline{g}^{\mu_{l}\mu_{m}}\partial_{\mu_{m}}\overline{g}_{\phi_{i}\phi_{j}}\left(\overline{\nabla}_{\mu_{l}}\overline{\nabla}_{0}h_{\phi_{k}r}+\overline{\nabla}_{\phi_{k}}\overline{\nabla}_{r}h_{0\mu_{l}}-\overline{\nabla}_{\mu_{l}}\overline{\nabla}_{r}h_{\phi_{k}0}-\overline{\nabla}_{\phi_{k}}\overline{\nabla}_{0}h_{\mu_{l}r}\right)\nonumber \\
 &  & ~~~~~~~~~~~~~~~~~~~~=-\frac{M}{2W}r^{2-D}(D-1)\overline{g}^{\phi_{j}\phi_{k}}\overline{g}^{\mu_{l}\mu_{m}}\partial_{\mu_{m}}\overline{g}_{\phi_{i}\phi_{j}}(2W\partial_{\mu_{l}}k_{\phi_{k}}+k_{\phi_{k}}\partial_{\mu_{l}}W).
\end{eqnarray}
Remember the equation (\ref{identity-1}), which reads $\overline{g}^{\mu_{k}\mu_{l}}\partial_{\mu_{l}}\overline{g}_{\phi_{i}\phi_{j}}\partial_{\mu_{k}}W=0$.
So, the last term in the last expression vanishes. After a straightforward
calculation, one can show the vanishing of the first one too. Then,
the last line in (\ref{gggg-1}) has no contribution to the angular
momentum. Inserting the results, we find
\begin{equation}
\left(R^{r0}\thinspace_{\beta\sigma}\right)^{\left(1\right)}\bar{\nabla}^{\beta}\bar{\xi}^{\sigma}=-2M\Lambda r^{2-D}(D-3)(D-1)k_{\phi_{i}},
\end{equation}
where we have $k_{\phi_{i}}=-\frac{a_{i}\mu_{i}^{2}}{1+\Lambda a_{i}^{2}}$,
from the equation (\ref{k_mu's}). So then, the integrand yields 
\begin{equation}
\left(R^{r0}\thinspace_{\beta\sigma}\right)^{\left(1\right)}\bar{\nabla}^{\beta}\bar{\xi}^{\sigma}=2M\Lambda r^{2-D}(D-3)(D-1)\frac{a_{i}\mu_{i}^{2}}{1+\Lambda a_{i}^{2}},
\end{equation}
and the angular momentum becomes
\begin{equation}
J_{i}=\frac{M(D-1)!!}{4}\int_{-1}^{1}\prod_{j=1}^{N}\frac{\mu_{j}d\mu_{j}}{1+\Lambda a_{j}^{2}}\thinspace\frac{a_{i}\mu_{i}^{2}}{(1+\Lambda a_{i}^{2})\sqrt{1-\sum_{k=1}^{N}\mu_{k}^{2}}}.
\end{equation}
Using the definition of $\Xi$, given in (\ref{=00005CXi-1}), one
can express the angular momentum as
\begin{equation}
J_{i}=\frac{M(D-1)!!}{4\Xi}\frac{a_{i}}{(1+\Lambda a_{i}^{2})}\int_{-1}^{1}\prod_{j=1}^{N}\frac{\mu_{j}d\mu_{j}}{\sqrt{1-\sum_{k=1}^{N}\mu_{k}^{2}}}\mu_{i}^{2}.
\end{equation}
From equation (\ref{integral3}) in Appendix B, the integral reads
\begin{equation}
\int_{-1}^{1}\prod_{j=1}^{N}\frac{\mu_{j}d\mu_{j}}{\sqrt{1-\sum_{k=1}^{N}\mu_{k}^{2}}}\mu_{i}^{2}=\frac{4}{(2N+1)!!}=\frac{4}{(D-1)!!}.
\end{equation}
One ends up with the angular momentum of the even dimensional Kerr-
AdS black holes corresponding to the Killing vector $\xi_{(i)}^{\mu}=\left(0,...,0,1_{i},0,...,0\right)$
as
\begin{equation}
J_{i}=\frac{Ma_{i}}{\Xi\Xi_{i}},
\end{equation}
 in cosmological Einstein's gravity. Note that, in odd spacetime dimensions
one also ends up with the same expression.

\end{document}